\documentclass[pra,twocolumn,tbtags,superscriptaddress,floatfix]{revtex4}
\usepackage{amssymb}
\usepackage{graphicx,amsmath}
\usepackage{bm}
\usepackage{times}
\DeclareGraphicsExtensions{.pdf,.png,.jpg,.eps}
\begin{document}
\title{Excitation of the collective states in a three- qubit system}

\begin{abstract}
In the present paper, we have proposed the experimentally
achievable method for the characterization of the collective
states of qubits in a linear chain. We study temporal dynamics
of absorption of a single-photon pulse by three interacting qubits
embedded in a one-dimensional waveguide. Numerical simulations
were performed for a Gaussian-shaped pulse with different
frequency detunings and interaction parameters between qubits. The
dynamic behavior of the excitation probability for each qubit is
investigated. It was shown that the maximum probability amplitudes
of excitation of qubits are reached when the frequency of external
excitation coincides with the frequency of excitation of the
a corresponding eigenstate of the system. In this case, the
the magnitude of the probability amplitude of each qubit in the chain
unambiguously correlates with the contribution of this qubit to
the corresponding collective state of the system, and the decay of
these amplitudes are determined by the resonance width arising from
the interaction of the qubit with the photon field of the
waveguide. Therefore, we show that the pulsed harmonic probe can
be used for the characterization of the energies, widths, and the
wavefunctions of the collective states in a one-dimensional qubit
chain.

%the rate of spontaneous emission $\Gamma$.

\end{abstract}

%\pacs{42.50 Ct,~ 42.50.Dv,~42.50.Pq}

\date{\today }
\author{Ya. S. Greenberg}\email{greenbergy@risp.ru}
\affiliation{Novosibirsk State Technical University, Novosibirsk,
Russia}

\author{A. A. Shtygashev}
\affiliation{Novosibirsk State Technical University, Novosibirsk,
Russia}

%\classification{85.25.Dq,~ 85.25.Cp,~ 85.25.Hv,~ 84.40.Az}

 \maketitle

 \section{Introduction}\label{Intr}
Quantum bits (qubits) are at the heart of quantum information
processing schemes. Currently, solid-state qubits, and in
particular the superconducting ones, seem to satisfy the
requirements for being the building blocks of viable quantum
computers, since they exhibit relatively long coherence times,
extremely low dissipation, and scalability. Furthermore, the
coupling between qubits has successfully been achieved that was
followed by the construction of multiple-qubit logic gates and the
implementation of several algorithms. Most of the information
protocols in qubit systems are based on a train of recording and
readout pulses. Mainly, the investigations are restricted to the
pulsed excitation of single-qubit \cite{Wang11, Stob09, Raph10,
Chen11, Dom02, Der13} or two-qubit systems \cite{Huang13, Der14,
Green18}. However, the existing quantum processors consist of at
least several tens of qubits \cite{list}. Therefore, the study of
the pulse excitation of multiqubit structures is of certain
interest. As shown in \cite{Liao15}, the behavior of multiqubit
structures under pulsed excitation has important features due to
the interaction of a photon with collective multiparticle states.
This interaction leads to such interesting physical effects as
photon blockade, Fano interference, quantum entanglement, and
superradiation radiation. In the present work, we investigate the
dynamic behavior under pulsed excitation of a linear chain
consisting of three qubits interacting with the photon field in a
one-dimensional waveguide. In principle, our method can be
extended to a linear chain of an arbitrary number of qubits. Here
we have focused our study on three-qubit chain because for the
energy spectrum of this system a simple analytical solution can be
obtained. This will allow us to attribute a clear physical meaning
to certain aspects of the dynamic behavior of the qubit excitation
amplitudes. In contrast to \cite{Liao15}, where a chain of real
atoms was studied, here we consider superconducting qubits, which,
unlike real atoms, have a technological spread in their parameters
(for example, the excitation energies of qubits differ in
principle from each other). Besides, the excitation energy of each
qubit can be individually tuned through external circuits. Another
difference is that we take into account the direct interaction
between qubits -the interaction of the Ising type between the
nearest neighbors. This interaction leads to the formation of
collective quasistationary states, the width of which is
determined by the interaction of each qubit with the photon field
of the waveguide. A numerical simulation was performed for a
Gaussian-shaped packet with different parameters of frequency
detuning and interaction between qubits. The dynamic behavior of
the excitation probability of each qubit is investigated. It is
shown that pulsed excitation makes it possible to identify
collective states of the system. The magnitude of the excitation
amplitude of each qubit in the chain is uniquely correlated with
the contribution of this qubit to the corresponding stationary
state of the system. The damping of these amplitudes is determined
by the resonance widths of the quasistationary states. The article
is organized as follows. In the first section, we consider a
linear chain of three qubits that interact with each other
according to the nearest neighbor Ising model. The wave functions
and the energy spectrum of this system have been found.  The
dependence of the parameters of stationary states on the degree of
non-identity of qubits has also been investigated. In the second
section, the effective Hamiltonian of a 3-qubit system is
investigated with account for spontaneous emission into the
waveguide. In the third section, the Wigner-Weisskopf
approximation, and a single-photon basis were used to obtain the
differential equations for the amplitudes of excitation of
individual qubits. The fourth section presents the results of
numerical simulations of the excitation amplitudes of individual
qubits under external excitation. It was shown that the magnitude
of the excitation amplitude of each qubit in the chain uniquely
correlates with the contribution of this qubit to the
corresponding stationary state of the system, and the damping of
these amplitudes is determined by the resonance width arising due
to the interaction of the qubit with the photon field of the
waveguide.

\section{Three interacting qubits. Wave functions and the energy
spectrum.}
We consider a linear chain of three equally spaced qubits which
are located at the points $x_1=-d, x_2=0, x_3=+d$. Every qubit
can be either in the excited, $ \left| {e } \right\rangle$ or the ground state $ \left| {g } \right\rangle$. The
Hamiltonian which accounts for the interaction between nearest
neighbor qubits is (we use units where $\hbar=1$ throughout this
paper):

\begin{equation}\label{H0}
H_0  = \frac{1}{2}\sum\limits_{n = 1}^3 {\left( {1 + \sigma
_z^{(n)} } \right)\Omega _n }  - J(\sigma _1^ +  \sigma _2  +
\sigma _2^ +  \sigma _1  + \sigma _3^ +  \sigma _2  + \sigma _2^ +
\sigma _3 )
\end{equation}
where $\Omega_n$-qubit excitation frequency, $J$-interqubit
coupling, $\sigma _n^ +   = \left| {e_n } \right\rangle
\left\langle {g_n } \right|,\,\sigma _n^{}  = \left| {g_n }
\right\rangle \left\langle {e_n } \right|$ are raising and
lowering Pauli operators, and $\sigma _z^{(n)} \left| {e_n }
\right\rangle = \left| {e_n } \right\rangle ,\,\sigma _z^{(n)}
\left| {g_n } \right\rangle  =  - \left| {g_n } \right\rangle $.
Here we assume that $J$ is not a photon mediated coupling. In
superconducting circuits with on-chip embedded qubits, the
interqubit coupling $J$ is controlled technologically, so that the
coupling between, say, first and third qubit may be absent no
matter how close they are in real space. Below we consider single
photon approximation with the only one qubit in the chain being
excited. Therefore, we will limit Hilbert space to three vector
states:

\begin{equation}\label{set1}
\left| 1 \right\rangle  = \left| {e_1 g_2 g_3 } \right\rangle,
\left| 2 \right\rangle  = \left| {g_1 e_2 g_3 } \right\rangle,
\left| 3 \right\rangle  = \left| {g_1 g_2 e_3 } \right\rangle
\end{equation}

The wave function is taken as a superposition of the vector states
(\ref{set1}):

\begin{equation}\label{Psi}
\Psi _i  = c_1^{(i)} \left| 1 \right\rangle  + c_2^{(i)} \left| 2
\right\rangle  + c_3^{(i)} \left| 3 \right\rangle ,\quad (i =
1,2,3)
\end{equation}

From Schrodinger equation, $H_0\Psi=E\Psi$ we obtain a linear
system which allows us to find the energies and the superposition
coefficients $c_i$ of our system:

\begin{equation}\label{matrix1}
    \left( {\begin{array}{*{20}c}
   {\Omega _1  - E} & { - J} & 0  \\
   { - J} & {\Omega _2  - E} & { - J}  \\
   0 & { - J} & {\Omega _3  - E}  \\
\end{array}} \right)\left( {\begin{array}{*{20}c}
   {c_1 }  \\
   {c_2 }  \\
   {c_3 }  \\
\end{array}} \right) = 0
\end{equation}
From (\ref{matrix1}) we obtain the equation for the energies:

\begin{equation}\label{En1}
    E^3  - a_2 E^2  - a_1 E - a_0  = 0
\end{equation}
where $ a_2  = \Omega _1  + \Omega _2  + \Omega _3$, $ a_1  = 2J^2
- \Omega _3 \Omega _2  - \Omega _3 \Omega _1  - \Omega _2 \Omega
_1 $, $ a_0  = \Omega _1 \Omega _2 \Omega _3  - (\Omega _1  +
\Omega _3 )J^2 $. If all qubits are identical ($ \Omega _1  =
\Omega _2  = \Omega _3  = \Omega $ ) we obtain from (\ref{En1})
the energies of the system:

\begin{equation}\label{En2}
\begin{array}{l}
 E_1  = \Omega  - \sqrt 2 J \\
 E_2  = \Omega  \\
 E_3  = \Omega  + \sqrt 2 J \\
 \end{array}
 \end{equation}

 The superposition coefficients $c_i$ are being calculated from (\ref{matrix1})
 taking into account the normalization:

 \begin{equation}\label{7}
    |c_1^{(i)} |^2  + |c_2^{(i)} |^2  + |c_3^{(i)} |^2  = 1,\quad (i = 1,2,3)
\end{equation}

Finally, the wave functions are as follows. For the lowest energy
$E_1  = \Omega  - \sqrt 2 J $

\begin{equation}\label{8a}
\Psi _1  = \frac{1}{2}\left| 1 \right\rangle  + \frac{{\sqrt 2
}}{2}\left| 2 \right\rangle  + \frac{1}{2}\left| 3 \right\rangle
\end{equation}

For $E=\Omega$
\begin{equation}\label{8b}
\Psi _2  = \frac{1}{{\sqrt 2 }}\left| 1 \right\rangle  + 0\left| 2
\right\rangle  - \frac{1}{{\sqrt 2 }}\left| 3 \right\rangle
\end{equation}
and for the highest energy $E_1  = \Omega  + \sqrt 2 J $

\begin{equation}\label{8c}
\Psi _3  = \frac{1}{2}\left| 1 \right\rangle  - \frac{{\sqrt 2
}}{2}\left| 2 \right\rangle  + \frac{1}{2}\left| 3 \right\rangle
\end{equation}

It is noteworthy that for identical qubits the superposition
coefficients $c_i$ do not depend on the coupling parameter $J$.
The wave functions (\ref{8a}, \ref{8b}, \ref{8c}) are the
collective states of a three- qubit chain which is given by the
Hamiltonian (\ref{H0}). Unlike the real atoms, superconducting
qubits are intrinsically not identical due to technological
scattering of their parameters. The excitation energy of every
qubit in a chain can moreover be adjusted to any value by an
external circuit. Below, we consider the situation when the
excitation frequency of one of the qubit is different from that of
the other two qubits. Therefore, we take the first and the third
qubit as identical ($\Omega _1  = \Omega _3  = \Omega $ ), while
the excitation frequency of the second qubit is  $\Omega_2$. A
direct calculation of the matrix determinant (\ref{matrix1})
yields the following result:

\begin{equation}\label{9a}
E_1  = \Omega  + \frac{\Delta }{2} - \frac{1}{2}\sqrt {\Delta ^2 +
8J^2 }
\end{equation}
\begin{equation}\label{9b}
    E_2  = \Omega
\end{equation}
\begin{equation}\label{9c}
E_3  = \Omega  + \frac{\Delta }{2} + \frac{1}{2}\sqrt {\Delta ^2 +
8J^2 }
\end{equation}
where $\Delta=\Omega_2-\Omega$ .

For eigen energies $E_1$ and $E_3$ the superposition coefficients
are as follows:
\begin{equation}\label{10a}
\begin{array}{l}
 c_1^{(1)}  = c_3^{(1)}  = \frac{J}{{\sqrt {2J^2  + 0.25\left( {\sqrt {\Delta ^2  + 8J^2 }  - \Delta } \right)^2 } }};\quad
 \\\\
 c_1^{(3)}  = c_3^{(3)}  = \frac{J}{{\sqrt {2J^2  + 0.25\left( {\sqrt {\Delta ^2  + 8J^2 }  + \Delta } \right)^2 } }} \\
 \end{array}
\end{equation}
\begin{equation}\label{10b}
\begin{array}{l}
 c_2^{(1)}  = \frac{1}{2}\frac{{ - \Delta  + \sqrt {\Delta ^2  + 8J^2 } }}{{\sqrt {2J^2  + 0.25\left( {\sqrt {\Delta ^2  + 8J^2 }  - \Delta } \right)^2 } }};\quad
 \\\\
 c_2^{(3)}  =  - \frac{1}{2}\frac{{\Delta  + \sqrt {\Delta ^2  + 8J^2 } }}{{\sqrt {2J^2  + 0.25\left( {\sqrt {\Delta ^2  + 8J^2 }  + \Delta } \right)^2 } }} \\
 \end{array}
\end{equation}

\begin{figure}
  \includegraphics[width=8cm]{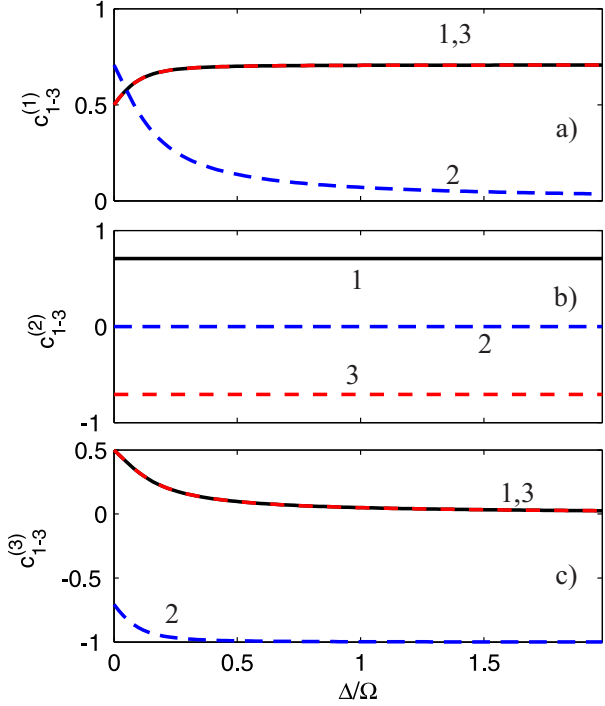}\\
  \caption{Dependence of superposition coefficients (\ref{10a},\ref{10b})
  on $\Delta/\Omega$ for $J/\Omega =0.05$. The numbers on the panels correspond to
  superposition coefficients: 1- $c_1^{(i)}$(black), 2-  $c_2^{(i)}$(blue),
  3- $c_3^{(i)}$(red).}\label{fig1}
\end{figure}

For the second stationary state $E_2$, superposition coefficients
coincide with those in (\ref{8b}). The dependence of superposition
coefficients on the parameter $\Delta/\Omega$ is shown in Fig.
\ref{fig1}. For the first energy state $\left| {E_1 } \right\rangle$ all three
coefficients $c_i^{(1)}$ become equal at the point $\Delta=J$
(panel $a)$ in Fig. \ref{fig1}).

As is follows from (\ref{10a},\ref{10b}), it results in the
formation at that point of a maximally entangled state $\left|
{E_1 } \right\rangle  = \frac{1}{{\sqrt 3 }}\left( {\left| {e_1
g_2 g_3 } \right\rangle  + \left| {g_1 e_2 g_3 } \right\rangle  +
\left| {g_1 g_2 e_3 } \right\rangle } \right)$. At this point, the
second state $\left| {E_2 } \right\rangle$ remains unaltered (\ref{8b}) while for the
third state we have $ \left| {E_3 } \right\rangle  =
\frac{1}{{\sqrt 6 }}\left| {e_1 g_2 g_3 } \right\rangle  -
\frac{2}{{\sqrt 6 }}\left| {g_1 e_2 g_3 } \right\rangle  +
\frac{1}{{\sqrt 6 }}\left| {g_1 g_2 e_3 } \right\rangle $ . It can
also be seen from ((\ref{10a},\ref{10b})) that as the detuning is
increased ($\Delta /J\gg 1$) the first state transforms to a
symmetrical entangled superposition $\left| {E_1 } \right\rangle =
\frac{1}{{\sqrt 2 }}\left( {\left| {e_1 g_2 g_3 } \right\rangle +
\left| {g_1 g_2 e_3 } \right\rangle } \right) $, the third state
is factorized $\left| {E_3 } \right\rangle  =  - \left| {g_1 e_2
g_3 } \right\rangle $ while the second state (\ref{8b}), remains
unaltered. Below, we show that under pulsed excitation of a qubit
chain, all these features emerge in the excitation spectrum of the
qubit probability amplitudes.

\section{Photon mediated interactions between qubits}

Spontaneous emission of qubits gives rise to the interqubit
coupling via the photon field in a waveguide. This interaction can
be described by the effective non-Hermitian Hamiltonian
\cite{Green15}:

\begin{widetext}
\begin{equation}\label{11}
\left\langle m \right|H_{eff} \left| n \right\rangle  = \left(
{\Omega _m  - i\Gamma _m } \right)\delta _{m,n}  - J_{n - 1}
\delta _{m,n - 1}  - J_n \delta _{m,n + 1}  - i\left( {\Gamma _m
\Gamma _n } \right)^{1/2} e^{ik\left| {d_{mn} } \right|}
\end{equation}
\end{widetext}
where $\Gamma_m$  is the rate of spontaneous emission of the m-th
qubit, $d_{mn}$ is a distance between the m-th and the n-th
qubits.

For three qubit chain the complex energies are derived by equating
the matrix determinant $ \left\langle {n|H_{eff}  - E|\left. m
\right\rangle } \right. $

\begin{widetext}
\begin{equation}\label{12a}
\left( {\begin{array}{*{20}c}
   {\Omega _1  - i\Gamma _1  - E} & { - i\sqrt {\Gamma _1 \Gamma _2 } e^{ikd}  - J} & { - i\sqrt {\Gamma _1 \Gamma _3 } e^{ik2d} }  \\
   { - i\sqrt {\Gamma _1 \Gamma _2 } e^{ikd}  - J} & {\Omega _2  - i\Gamma _2  - E} & { - i\sqrt {\Gamma _3 \Gamma _2 } e^{ikd}  - J}  \\
   { - i\sqrt {\Gamma _3 \Gamma _1 } e^{ik2d} } & { - i\sqrt {\Gamma _3 \Gamma _2 } e^{ikd}  - J} & {\Omega _3  - i\Gamma _3  - E}  \\
\end{array}} \right)
\end{equation}
\end{widetext}
to zero.

For identical qubits ($ \Omega_i=\Omega ,\Gamma_i=\Gamma$ ) and
long wavelength limit ($kd\ll1$) we obtain from (\ref{12a}) the
following matrix
\begin{equation}\label{12b}
\left( {\begin{array}{*{20}c}
   {\Omega  - i\Gamma  - E} & { - i\Gamma  - J} & { - i\Gamma }  \\
   { - i\Gamma  - J} & {\Omega  - i\Gamma  - E} & { - i\Gamma  - J}  \\
   { - i\Gamma } & { - i\Gamma  - J} & {\Omega  - i\Gamma  - E}  \\
\end{array}} \right)
\end{equation}
As is well known \cite{Green15,Volya03}, the energy spectrum of
the effective Hamiltonian matrix (\ref{12b}) with $J=0$ has a
simple structure. There exists the only superradiant
non-stationary state with the energy $E = \Omega  - i3\Gamma $ and
two degenerate stable states with the energy $E=\Omega$. If $J$ is
different from zero the energy spectrum found from (\ref{12b}) is
as follows:
\begin{equation}\label{13}
\begin{array}{l}
 E_1  = \Omega  - i\frac{3}{2}\Gamma  - \sqrt { - \frac{9}{4}\Gamma _{}^2  + 2J^2  + 4iJ\Gamma }  \\
 E_2  = \Omega  \\
 E_3  = \Omega  - i\frac{3}{2}\Gamma  + \sqrt { - \frac{9}{4}\Gamma _{}^2  + 2J^2  + 4iJ\Gamma }  \\
 \end{array}
\end{equation}
Here, there are one stable and two unstable states. If $J$ tends
to zero we obtain from (\ref{13}) two stable degenerate states and
one unstable superradiant state. If the excitation frequency of a
central qubit is defferent from that of other two qubits in a
chain we obtain from matrix determinant (\ref{12b}):

\begin{widetext}
\begin{equation}\label{14}
\begin{array}{l}
 E_1  = \Omega  + \frac{1}{2}\Delta  - i\frac{3}{2}\Gamma  - \sqrt {\frac{1}{4}\left( {\Delta  - 3i\Gamma } \right)^2  + 2i\Gamma \Delta  + 2J^2  + 4iJ\Gamma }  \\
 E_2  = \Omega  \\
 E_3  = \Omega  + \frac{1}{2}\Delta  - i\frac{3}{2}\Gamma  + \sqrt {\frac{1}{4}\left( {\Delta  - 3i\Gamma } \right)^2  + 2i\Gamma \Delta  + 2J^2  + 4iJ\Gamma }  \\
 \end{array}
\end{equation}
\end{widetext}
where $\Delta=\Omega_2-\Omega$.

If $\Delta=J$ we obtain from (\ref{14}):
\begin{equation}\label{15}
\begin{array}{l}
 E_1  = \Omega  - \Delta  - i3\Gamma  \\
 E_2  = \Omega  \\
 E_3  = \Omega  + 2\Delta  \\
 \end{array}
\end{equation}

\section{Three qubit chain under the influence of pulsed excitation}

Here we consider a time dependence of the excitation probability
for every qubit, $\beta_n(t)$ in the chain subjected to pulsed
excitation. We start with the Hamiltonian, which describes the
interaction of qubits with the photon field in a waveguide:

\begin{widetext}
\begin{equation}\label{16}
H = H_0  + \sum\limits_k { \omega _k a_k^ +  a_k }  +
\sum\limits_{n = 1}^3 {\sum\limits_k {\left( {\left( {g_k^{(n)}
e^{ - ikx_n } \sigma _ - ^{(n)} } \right)a_k^ +   + h.c.} \right)}
}
\end{equation}
\end{widetext}
where $H_0$ is given in (\ref{H0}).

The quantities $g_k^{(i)} ,\;(i = 1,2,3) $  in (\ref{16}) describe
the qubit interaction with the photon field in a waveguide:
\begin{equation}\label{17}
g_k^{(i)}  = \sqrt {\frac{{\omega _k^{} D_i ^2 }}{{2\hbar
\varepsilon _0 V}}}
\end{equation}
where $D_i$ is a dipole moment of the $i-$th qubit, $V$ is the
effective volume of the  photon-qubit interaction.

We will consider only single-photon states when one photon is
present in the system and qubits are in the ground state, or one
of the qubits is excited, and there are no photons in the system.
Following this, we write the state vector as follows:

\begin{equation}\label{18}
\left| \Psi  \right\rangle  = \sum\limits_{n = 1}^3 {\beta _n
(t)e^{ - i\Omega _n t} } \left| n \right\rangle  + \sum\limits_k
{\gamma _k (t)e^{ - i\omega _k t} } \left| {G,k}
\right\rangle
\end{equation}
where $ \left| 1 \right\rangle  = \left| {e_1 g_2 g_3 0_k }
\right\rangle $, $ \left| 2 \right\rangle  = \left| {g_1 e_2 g_3
0_k } \right\rangle $, $ \left| 3 \right\rangle  = \left| {g_1 g_2
e_3 0_k } \right\rangle $, $ \left| {G,k} \right\rangle  = \left|
{g_1 g_2 g_3 ,1_k } \right\rangle $.

The equations for the amplitudes $ \gamma _k (t),\,\beta _1
(t),\,\beta _2 (t),\,\beta _3 (t) $  are derived from Schrodinger
equation $id\left| \Psi  \right\rangle /dt = H\left| \Psi
\right\rangle $.

\begin{widetext}
\begin{equation}\label{19a}
\begin{array}{l}
 \frac{{d\beta _1 }}{{dt}} =  - i\sum\limits_k {g_k^{*(1)} e^{ikx_1 } e^{ - i(\omega _k  - \Omega _1 )t} } \tilde \gamma _k (0) - \sum\limits_k {|g_k^{(1)} |^2 } \int\limits_0^t {\beta _1 (t')e^{ - i(\omega _k  - \Omega _1 )(t - t')} dt'}  \\
 \quad \quad \quad \quad \quad  - \sum\limits_k {g_k^{*(1)} g_k^{(2)} e^{ - ik(x_2  - x_1 )} e^{i(\Omega _1  - \Omega _2 )t} } \int\limits_0^t {\beta _2 (t')e^{ - i(\omega _k  - \Omega _2 )(t - t')} dt'}  \\
 \quad \quad \quad \quad \quad  - \sum\limits_k {g_k^{*(1)} g_k^{(3)} e^{ - ik(x_3  - x_1 )} e^{i(\Omega _1  - \Omega _3 )t} } \int\limits_0^t {\beta _3 (t')e^{ - i(\omega _k  - \Omega _3 )(t - t')} dt'}  \\
 \quad \quad \quad \quad \quad  + iJe^{i(\Omega _1  - \Omega _2 )t} \beta _2  \\
 \end{array}
\end{equation}

\begin{equation}\label{19b}
\begin{array}{l}
 \frac{{d\beta _2 }}{{dt}} =  - i\sum\limits_k {g_k^{*(2)} e^{ikx_2 } e^{ - i(\omega _k  - \Omega _2 )t} } \tilde \gamma _k (0) - \sum\limits_k {|g_k^{(2)} |^2 } \int\limits_0^t {\beta _2 (t')e^{ - i(\omega _k  - \Omega _2 )(t - t')} dt'}  \\
 \quad \quad \quad \quad \quad  - \sum\limits_k {g_k^{*(2)} g_k^{(1)} e^{ - ik(x_1  - x_2 )} e^{i(\Omega _2  - \Omega _1 )t} } \int\limits_0^t {\beta _1 (t')e^{ - i(\omega _k  - \Omega _1 )(t - t')} dt'}  \\
 \quad \quad \quad \quad \quad  - \sum\limits_k {g_k^{*(2)} g_k^{(3)} e^{ - ik(x_3  - x_2 )} e^{i(\Omega _2  - \Omega _3 )t} } \int\limits_0^t {\beta _3 (t')e^{ - i(\omega _k  - \Omega _3 )(t - t')} dt'}  \\
 \;{\rm{                   }} + iJe^{i(\Omega _2  - \Omega _1 )t} \beta _1  + iJe^{i(\Omega _2  - \Omega _3 )t} \beta _3  \\
 \end{array}
\end{equation}

\begin{equation}\label{19c}
\begin{array}{l}
 \frac{{d\beta _3 }}{{dt}} =  - i\sum\limits_k {g_k^{*(3)} e^{ikx_3 } e^{ - i(\omega _k  - \Omega _3 )t} } \tilde \gamma _k (0) - \sum\limits_k {|g_k^{(3)} |^2 } \int\limits_0^t {\beta _3 (t')e^{ - i(\omega _k  - \Omega _3 )(t - t')} dt'}  \\
 \quad \quad \quad \quad \quad  - \sum\limits_k {g_k^{*(3)} g_k^{(1)} e^{ - ik(x_1  - x_3 )} e^{i(\Omega _3  - \Omega _1 )t} } \int\limits_0^t {\beta _1 (t')e^{ - i(\omega _k  - \Omega _1 )(t - t')} dt'}  \\
 \quad \quad \quad \quad \quad  - \sum\limits_k {g_k^{*(3)} g_k^{(2)} e^{ - ik(x_2  - x_3 )} e^{i(\Omega _3  - \Omega _2 )t} } \int\limits_0^t {\beta _3 (t')e^{ - i(\omega _k  - \Omega _3 )(t - t')} dt'}  \\
 {\rm{                      }} + iJe^{i(\Omega _3  - \Omega _2 )t} \beta _2  \\
 \end{array}
\end{equation}

\begin{equation}\label{20}
\begin{array}{l}
 \gamma _k (t) = \tilde \gamma _k (0) - ig_k^{(1)} e^{ - ikx_1 }
  \int\limits_0^t {\beta _1 (t')e^{i(\omega _k  - \Omega _1 )t'} dt'}  \\
 \quad \quad \quad  - ig_k^{(2)} e^{ - ikx_2 } \int\limits_0^t
  {\beta _2 (t')e^{i(\omega _k  - \Omega _2 )t'} dt'}  - ig_k^{(3)} e^{ - ikx_3 }
  \int\limits_0^t {\beta _3 (t')e^{i(\omega _k  - \Omega _3 )t'} dt'}  \\
 \end{array}
\end{equation}
\end{widetext}
Here $ \tilde \gamma _k (0)=\sqrt{2/L}\gamma_k(0) $ where $L$ is a
waveguide length, and $\gamma_k(0)$ is the initial Gaussian
envelope:

\begin{equation}\label{21}
\gamma _k (0) =  \left( {\frac{2}{{\pi \Delta_k ^2 }}}
\right)^{1/4} \exp \left( {i(k - k_s )x_0  - \frac{{(k - k_s )^2
}}{{\Delta_k ^2 }}} \right)
\end{equation}
 where $\Delta_k$ is a spectral width of
the packet in $k$ space, which is related to a spatial width of
the packet: $\sigma=\sqrt{2}/\Delta_k$ , $-x_0$ is the position of
the maximum of envelope curve on $x$-axis at the initial moment of
time, $k_s =\omega_s/v_g$ is the center of the wave packet in the
$k$ space, $\omega_s$ is the frequency of the center of the photon
pulse, $v_g$ is the group velocity of the wave in a waveguide.

In the framework of the single-photon approximation, the system of
equations (\ref{19a}, \ref{19b}, \ref{19c}) is accurate. A further
simplification of this system is associated with the
Wigner-Weisskopf approximation, which allows us to express the
photon-qubit interaction couplings $g_k^{(i)} $  in terms of the
rate of spontaneous decay, $\Gamma_i$ of the $i$-th qubit into the
waveguide,
 (see the appendix).

\begin{equation}\label{Gamma}
    \Gamma _i  = 4L\left| {g^{(i)} (\Omega )} \right|^2
/{\rm{v}}_g
\end{equation}

As is shown in the appendix, in the Wigner-Weisskopf
approximation, equations (\ref{19a}, \ref{19b}, \ref{19c}) for the
excitation amplitudes , $\beta_n(t)$ can be written in the
following form:

\begin{widetext}
\begin{equation}\label{22a}
\begin{array}{l}
 \frac{{d\beta _1 }}{{dt}} =  - i\sqrt {\frac{{\Gamma _1 v_g }}{{4\pi }}} \left( {\frac{{\omega _s }}{{\Omega _1 }}} \right)^{1/2} \exp \left( {i\Omega _1 t} \right)f(k_s ,x_1 ,t) - \frac{{\Gamma _1 }}{2}\beta _1 (t) + iJe^{i(\Omega _1  - \Omega _2 )t} \beta _2 (t)
 \\\\
  - \frac{{\sqrt {\Gamma _1 \Gamma _2 } }}{2}\sqrt {\frac{{\Omega _2 }}{{\Omega _1 }}} e^{ - ik_2 d} e^{i(\Omega _1  - \Omega _2 )t} \beta _2 (t) - \frac{{\sqrt {\Gamma _1 \Gamma _3 } }}{2}\sqrt {\frac{{\Omega _3 }}{{\Omega _1 }}} e^{ - ik_3 2d} e^{i(\Omega _1  - \Omega _3 )t} \beta _3 (t) \\
 \end{array}
\end{equation}

\begin{equation}\label{22b}
\begin{array}{l}
 \frac{{d\beta _2 }}{{dt}} =  - i\sqrt {\frac{{\Gamma _2 v_g }}{{4\pi }}} \left( {\frac{{\omega _s }}{{\Omega _2 }}} \right)^{1/2} \exp \left( {i\Omega _2 t} \right)f(k_s ,x_2 ,t) - \frac{{\Gamma _2 }}{2}\beta _2 (t)
 \\\\
  - \frac{{\sqrt {\Gamma _2 \Gamma _1 } }}{2}\sqrt {\frac{{\Omega _1 }}{{\Omega _2 }}} e^{ik_1 d} e^{i(\Omega _2  - \Omega _1 )t} \beta _1 (t) - \frac{{\sqrt {\Gamma _2 \Gamma _3 } }}{2}\sqrt {\frac{{\Omega _3 }}{{\Omega _2 }}} e^{ - ik_3 d} e^{i(\Omega _2  - \Omega _3 )t} \beta _3 (t)
  \\\\
 \quad \quad  + iJe^{i(\Omega _2  - \Omega _1 )t} \beta _1 (t) + iJe^{i(\Omega _2  - \Omega _3 )t} \beta _3 (t) \\
 \end{array}
\end{equation}

\begin{equation}\label{22c}
\begin{array}{l}
 \frac{{d\beta _2 }}{{dt}} =  - i\sqrt {\frac{{\Gamma _2 v_g }}{{4\pi }}} \left( {\frac{{\omega _s }}{{\Omega _2 }}} \right)^{1/2} \exp \left( {i\Omega _2 t} \right)f(k_s ,x_2 ,t) - \frac{{\Gamma _2 }}{2}\beta _2 (t)
 \\\\
  - \frac{{\sqrt {\Gamma _2 \Gamma _1 } }}{2}\sqrt {\frac{{\Omega _1 }}{{\Omega _2 }}} e^{ik_1 d} e^{i(\Omega _2  - \Omega _1 )t} \beta _1 (t) - \frac{{\sqrt {\Gamma _2 \Gamma _3 } }}{2}\sqrt {\frac{{\Omega _3 }}{{\Omega _2 }}} e^{ - ik_3 d} e^{i(\Omega _2  - \Omega _3 )t} \beta _3 (t)
  \\\\
 \quad \quad  + iJe^{i(\Omega _2  - \Omega _1 )t} \beta _1 (t) + iJe^{i(\Omega _2  - \Omega _3 )t} \beta _3 (t) \\
 \end{array}
\end{equation}

where the quantity $f(k_s,x,t)$ is the harmonic filled Gaussian
envelope in real space:
\begin{equation}\label{Gauss1}
f(k_s ,x,t) = \left( {2\pi \Delta _k^2 } \right)^{1/4} \exp \left(
{ik_s (x - \upsilon _g t) - \frac{{\Delta _k^2 }}{4}\left( {x_0  +
x - \upsilon _g t} \right)^2 } \right)
\end{equation}
\end{widetext}
It can be expressed in terms of Gaussian envelope  $\gamma_k(0)$
(\ref{21}) in the k space:
\begin{equation}\label{Gauss2}
f(k_s ,x,t) = \int\limits_{ - \infty }^\infty  {dk\gamma _k
(0)e^{ik(x - {\rm{v}}_g t)} }
\end{equation}

\section{Numerical calculations of the qubits excitation amplitudes,
$\beta_n(t)$}

Below, we present the results of numerical calculations of the
excitation amplitudes of a system of three qubits where the
interaction between the nearest neighbors is taken into account.
We solve the differential equations (\ref{22a}, \ref{22b},
\ref{22c}) with the initial conditions: $\beta_i(0)=0, (i=1,2,3)$,
with the harmonic filled Gaussian envelope (\ref{Gauss1}) at the
initial moment of time, $f(k_s,x,0)$. In this case, the width of
the initial Gaussian packet $\Delta_k$ in (\ref{21}) was chosen to
ensure a maximum of the excitation amplitude upon excitation of a
\emph{single} qubit. As was shown in \cite{Chen11} it can be
achieved if $\Delta_k=\Gamma/v_g$. For the values $\Gamma/2\pi=10$
MHz, $v_g=10^8 $ m/c we obtain $\Delta_k=0.21$ m$^{-1}$. The time
which takes for the center of the Gaussian envelope to reach the
first qubit in the chain was chosen to exceed the spontaneous
decay of the qubit excitation into the waveguide, $2/\Gamma$.
Therefore, we take $x_0=10v_g/\Gamma\approx 47.74$m. In all plots
the time is normalized to  $\tau=1/\Gamma= 1.59\times 10^{-8}$c.

\subsection{The excitation of identical qubits}
Figure 2 shows the spectroscopy calculated according to equations
(\ref{22a}, \ref{22b}, \ref{22c}) of the maximum values of the
excitation probabilities of each qubit depending on the excitation
frequency of the external signal $\omega_s$. The calculations were
carried out for identical qubits for $d = 1 $mm, $J/\Omega =0.05$,
$\Omega/2\pi =5$ GHz, $\Gamma/2\pi =10$ MHz.
\begin{figure}
  \includegraphics[width=8cm]{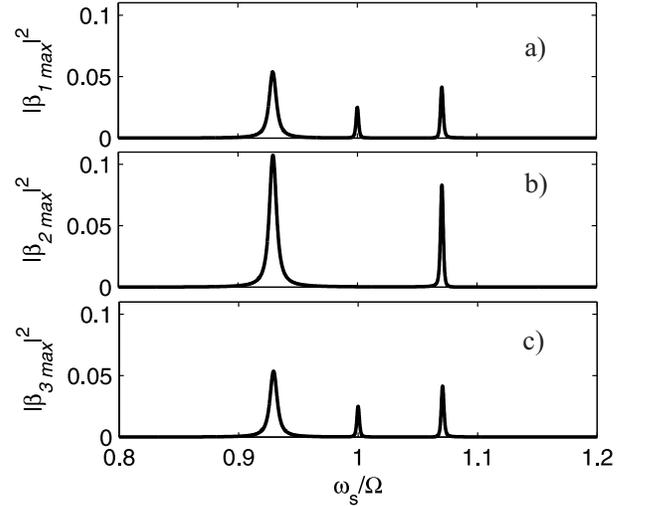}\\
  \caption{Spectroscopy of a three-qubit system. The dependence of the maximum
  excitation probability of the qubits on the single-photon probing
  frequency $\omega_s$. $d = 1 $mm, $J/\Omega =0.05$,
$\Omega/2\pi =5$ GHz, $\Gamma/2\pi =10$ MHz.}\label{fig2}
\end{figure}\\

For these values we obtain from the determinant (\ref{12a}) the
complex energies:

\begin{equation}\label{23}
\begin{array}{l}
 E_1 /\Omega  \approx 0.9293 - i5.82 \times 10^{ - 3}  \\
 E_2 /\Omega  \approx 1 \\
 E_3 /\Omega  \approx 1.0707 - i1.71 \times 10^{ - 4}  \\
 \end{array}
\end{equation}

Therefore, the first level has maximal width. The width of the
third level is less, while the width of the second level is
theoretically zero if inter qubit space, $d=0$. For $d=1$mm we
used in the calculations, the width of the second level is much
less than those of the other ones. These features are clearly seen
in Fig.\ref{fig2}. The peak positions and their widths are well
correlated with the real and imaginary parts of (\ref{23}).
Moreover, the peak heights are also well corresponded with the
squared values of the superposition coefficients $|c_n^{(i)}|^2$
in the collective wave functions (\ref{8a}, \ref{8b}, \ref{8c}).

For instance, if the probe frequency $\omega_s$ is equal to $E_1$,
the relative values of the peak heights (left peaks in
Fig.\ref{fig2}) are similar to those of superposition coefficients
in (\ref{8a}) with the amplitudes of the first and the third
qubits being equal, while the amplitude of the second qubit is
$\sqrt{2}$ times more. If the probe frequency is equal to $E_2$
(central peaks in Fig.\ref{fig2}), the probability amplitude for
the excitation of the second qubit (panel b) in Fig.\ref{fig2}) is
zero, which agrees with (8b).

\begin{figure}
  \includegraphics[width=8cm]{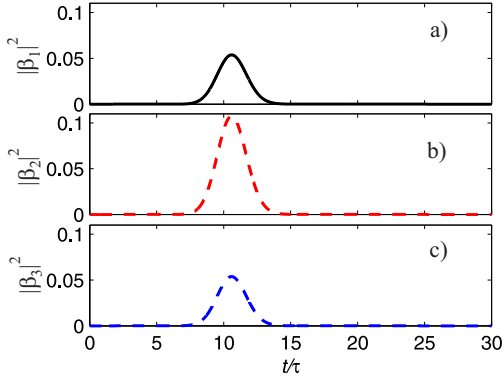}\\
  \caption{The time evolution of the qubits excitation probabilities
   $|\beta_n(t)|^2$, when the probe frequency is tuned to the first energy level
   $\omega_s=ReE_1=0.9293\Omega$ , $d=1$mm, $J/\Omega =0.05$,
   $\Omega/2\pi =5$GHz, $\Gamma/2\pi =10$ MHz. }\label{fig3}
\end{figure}

\begin{figure}
  \includegraphics[width=8cm]{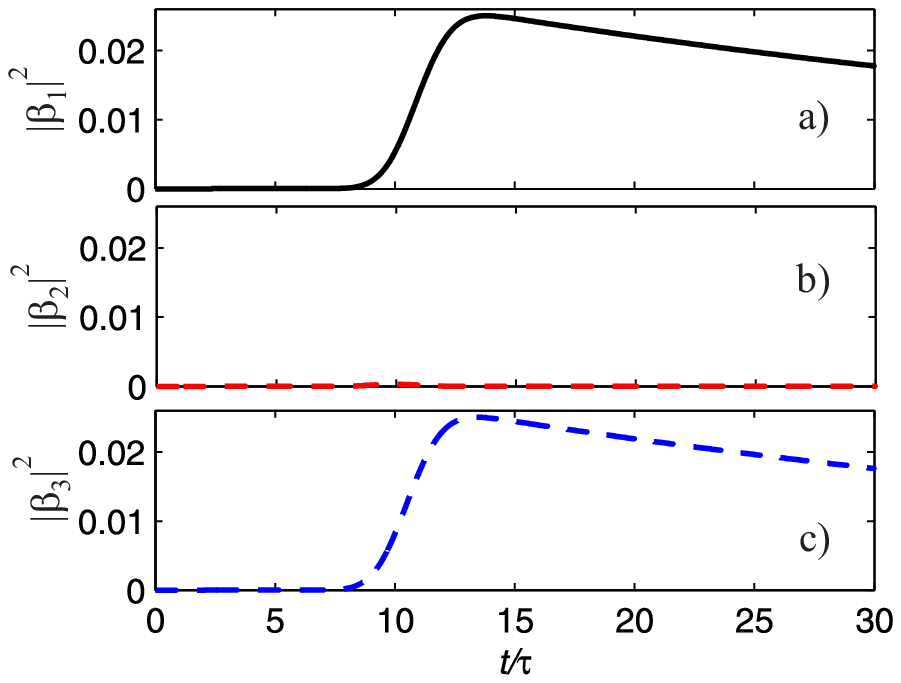}\\
 \caption{The time evolution of the qubits excitation probabilities
   $|\beta_n(t)|^2$, when the probe frequency is tuned to the second energy level
   $\omega_s=ReE_2=\Omega$ , $d=1$mm, $J/\Omega =0.05$,
   $\Omega/2\pi =5$GHz, $\Gamma/2\pi =10$ MHz. }\label{fig4}
\end{figure}

\begin{figure}
  \includegraphics[width=8cm]{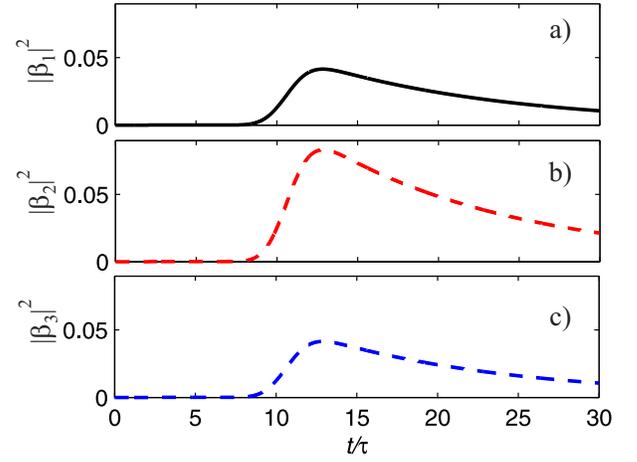}\\
 \caption{The time evolution of the qubits excitation probabilities
   $|\beta_n(t)|^2$, when the probe frequency is tuned to the third energy level
   $\omega_s=ReE_3=1.0707\Omega$ , $d=1$mm, $J/\Omega =0.05$,
   $\Omega/2\pi =5$GHz, $\Gamma/2\pi =10$ MHz. }\label{fig5}
\end{figure}

The time evolution of the qubits excitation probabilities
$|\beta_n(t)|^2$, when the probe frequency is tuned, respectively,
to first, second, and third energy levels (see (\ref{23}) is shown
in Figs.\ref{fig3},  \ref{fig4}, \ref{fig5}.

\begin{figure}
  \includegraphics[width=8 cm]{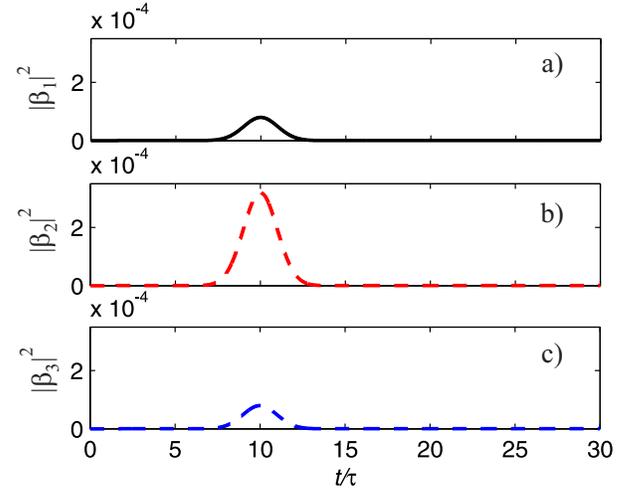}\\
  \caption{The time evolution of the qubits excitation probabilities
   $|\beta_n(t)|^2$, when the probe frequency
  is tuned to the second energy level  $\omega_s=ReE_2=\Omega$ , $d=0$, $J/\Omega =0.05$,
   $\Omega/2\pi =5$GHz, $\Gamma/2\pi =10$ MHz. }\label{fig6}
\end{figure}

As is seen in these figures, the relative value of the amplitudes
agrees with the contribution of a given qubit in the superposition
functions (\ref{8a}, \ref{8b}, \ref{8c}). For instance, the
excitation amplitude of the central qubit in Fig.\ref{fig4} is
equal to zero, because its contribution in a wavefunction
(\ref{8b}) is also equal to zero. The time dependence of the
amplitudes corresponds with the widths of relevant resonances in
(\ref{23}). For example, the first level $E_1$ has the maximum
width. Therefore, as is seen in Fig.\ref{fig3}, the excitation
amplitudes for every qubit rapidly decay. The width of the third
level $E_3$ is much less than that of the first one. If the probe
frequency is tuned to the frequency of the third level, it results
in a slow decay of the excitation amplitudes (see Fig.\ref{fig5}).
Much slower decay is observed if the probe frequency is tuned to
the second level $E_2$ (see Fig 4). As we pointed out before, the
width of the second level $E_2$  is only due to the non zero value
of d. For $d=1$ mm, which we used in the calculations, this width
is quite small resulting in a very slow decay of the amplitudes in
Fig.\ref{fig4}. We attribute this slow decay to solely the
interference effects due to the finite distance between qubits. If
we assume $d=0$ in equations (\ref{22a}, \ref{22b}, \ref{22c}),
the probability of the qubit excitation is greatly reduced (see
Fig.\ref{fig6}).

\subsection{Excitation of non identical qubits}

Below, we consider the excitation of the three-qubit chain when
the excitation frequency of one of the qubit is different from
that of the other two qubits. Therefore, we take the first and the
third qubit as identical ($\Omega_1=\Omega_3=\Omega$), while the
excitation frequency of the second qubit is $\Omega_2$. We also
assume all the rates of spontaneous emission as identical
($\Gamma_i=\Gamma$). First, we consider the case of large detuning
when $\Delta/\Omega=0.5$ ($ \Omega_2/\Omega=1.5$). The calculated
values of resonances and relevant superposition coefficients found
from the Hamiltonian matrix (\ref{12a}) for $J/\Omega=0.05$,
$\Gamma=10$ MHz, $d=1$ mm are presented in Table \ref{table}.
\begin{widetext}
\begin{center}
\begin{table}
\caption{The resonances and relevant superposition coefficients
for $J/\Omega =0.05, \Gamma/2\pi=10$ MHz, $d=1$ mm.}\label{table}
\begin{tabular}{ | l | l | l | l |}
\hline $E_j/\Omega$ & $c_1^{(j)}$ & $c_2^{(j)}$ &$c_3^{(j)}$\\
\hline
$0.9907\!-\!0.004683\it{i}$ & 0.7004 & 0.1370+0.00475\it{i} &0.7004\\
$0.9996\!-\!0.000437\it{i}$ & 0.7071 & 0.0000 &-0.7071 \\
$1.5097\!-\!0.001273\it{i}$ & 0.09686+0.00336\it{i} & -0.9906 &0.09686+0.00336\it{i}\\
\hline
\end{tabular}
\end{table}
\end{center}
\end{widetext}
As can be seen from this table, the values of the superposition
coefficients  correlate well with the asymptotic behavior of
expressions (\ref{10a}, \ref{10b}) for $\Delta/J \gg 1$.

\begin{figure}
  \includegraphics[width=8 cm]{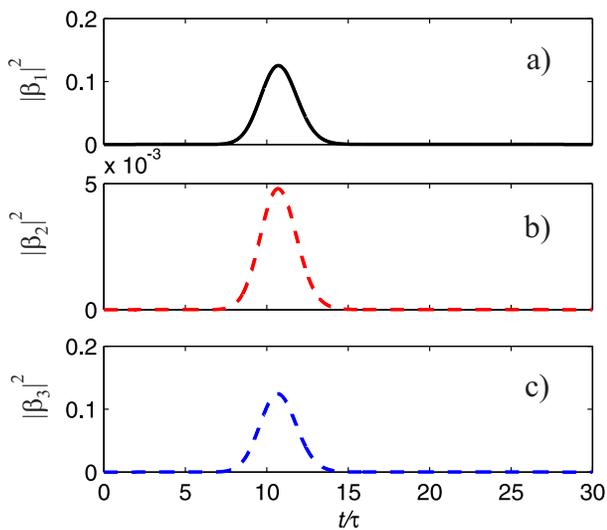}\\
  \caption{The time evolution of the qubits excitation probabilities
   $|\beta_n(t)|^2$, when the probe frequency
  is tuned to the first resonance   $\omega_s=$Re $E_1=0.9903\Omega$ , $d=1$ mm, $J/\Omega =0.05$,
   $\Omega/2\pi =5$GHz, $\Gamma/2\pi =10$ MHz, $\Delta /\Omega =0.5 ( \Omega_2/\Omega =1.5)$.}\label{fig7}
\end{figure}

The time dependence of the excitation amplitudes at the frequency
of the first resonance (Re $E_1/\Omega= 0.9907$)is shown in
Fig.\ref{fig7}.
The excitation amplitudes of the first and second qubits are the
same, and the central qubit is practically not excited. Since this
resonance has a relatively large width, the amplitudes decay
relatively quickly.

\begin{figure}
  \includegraphics[width=8 cm]{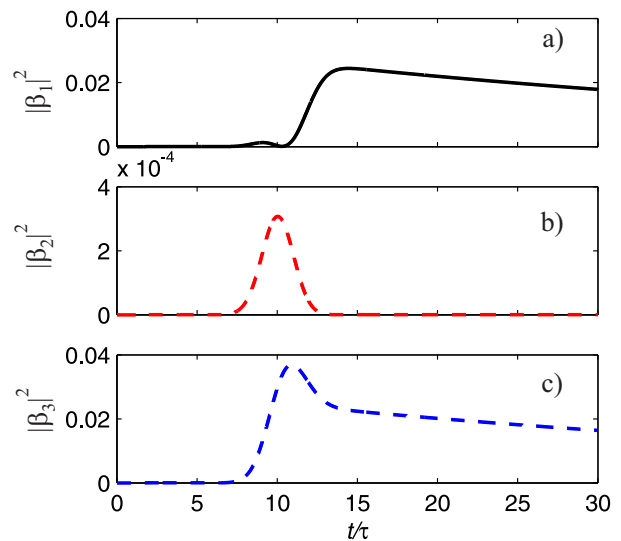}\\
  \caption{The time evolution of the qubits excitation probabilities
   $|\beta_n(t)|^2$, when the probe frequency
  is tuned to the second resonance   $\omega_s=$Re $E_2=0.9999\Omega$ , $d=1$ mm, $J/\Omega =0.05$,
   $\Omega/2\pi =5$GHz, $\Gamma/2\pi =10$ MHz, $\Delta /\Omega=0.5$
   $( \Omega_2/\Omega =1.5$}\label{fig8}
\end{figure}

When excited at the frequency of the second resonance (Re$
E_2/\Omega = 0.9996$), whose width is quite small, we see a
subradiant mode when the first and third qubits are excited
(Fig.\ref{fig8}), and the central qubit is practically not excited
(see Table\ref{table}).

\begin{figure}
  \includegraphics[width=8 cm]{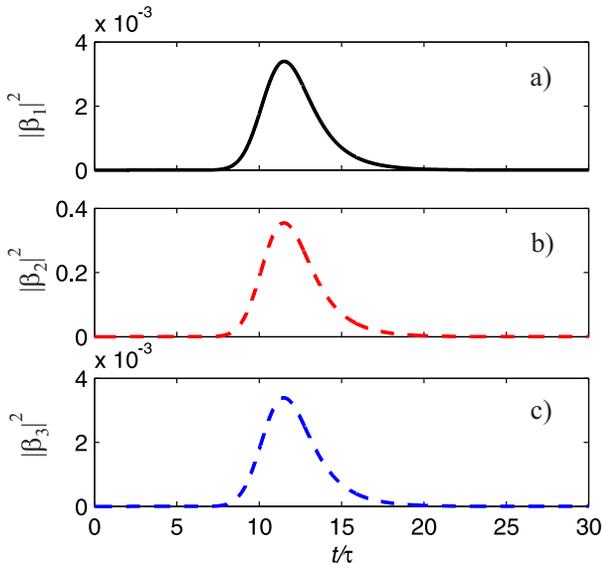}\\
  \caption{The time evolution of the qubits excitation probabilities
   $|\beta_n(t)|^2$, when the probe frequency
  is tuned to the third resonance   $\omega_s=$Re $E_3=1.5\Omega$ , $d=1$ mm, $J/\Omega =0.05$,
   $\Omega/2\pi =5$GHz, $\Gamma/2\pi =10$ MHz, $\Delta /\Omega=0.5$
   $( \Omega_2/\Omega =1.5$}\label{fig9}
\end{figure}

When excited at the frequency of the third resonance (Re$
E_3/\Omega= 1.51$), the second qubit is mainly excited
(Fig.\ref{fig9}), since the contributions of the first and third
qubits to the wave function of the third level are relatively
small (see the last line in Table\ref{table}).

\subsubsection{Excitation of non identical qubits with $\Delta=J$}

\begin{figure}
  \includegraphics[width=8 cm]{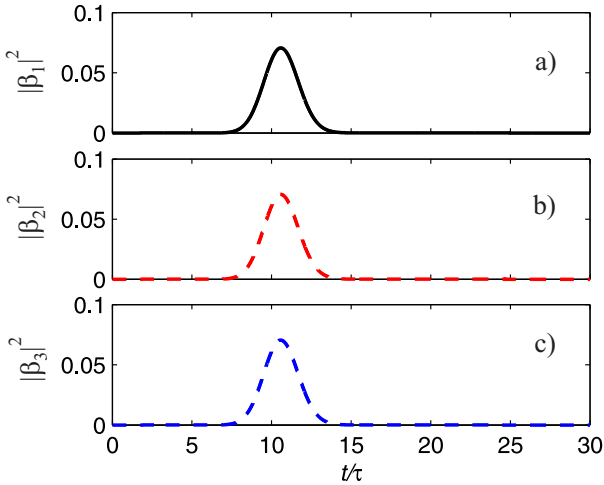}\\
  \caption{The time evolution of the qubits excitation probabilities
   $|\beta_n(t)|^2$, when the probe frequency
  is tuned to the first resonance   $\omega_s=$Re $E_1=0.95\Omega$ ,
  $d=1$ mm, $J/\Omega =0.05$,
   $\Omega/2\pi =5$GHz, $\Gamma/2\pi =10$ MHz, $\Delta /\Omega =0.05
   ( \Omega_2/\Omega =1.05)$.}\label{fig10}
\end{figure}

In the last part of this section we consider the dynamics of the
excitation of a three-qubit system for   $\Delta= J$. It
corresponds to the point of intersection of the graphs in panel a)
in Fig.\ref{fig1}. In this case, the state with the lowest energy
has a finite width (see first equation in (\ref{15}), and, as
indicated above, the corresponding wave function of the stationary
state has the maximum entanglement. The dynamics of the excitation
of this state is shown in Fig.\ref{fig10}. The amplitudes of the
excitation of qubits are the same and decay quickly enough.

\begin{figure}
  \includegraphics[width=8 cm]{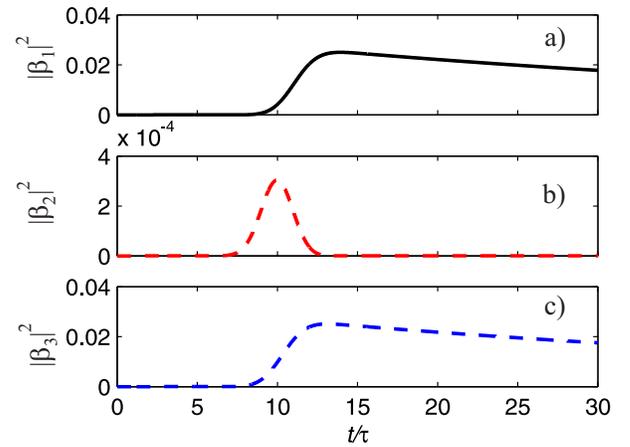}\\
  \caption{The time evolution of the qubits excitation probabilities
   $|\beta_n(t)|^2$, when the probe frequency
  is tuned to the second resonance   $\omega_s=$Re $E_2=1.0\Omega$ ,
  $d=1$ mm, $J/\Omega =0.05$,
   $\Omega/2\pi =5$GHz, $\Gamma/2\pi =10$ MHz, $\Delta /\Omega =0.05
   ( \Omega_2/\Omega =1.05)$.}\label{fig11}
\end{figure}

The second energy level of this system has practically no width
(second line in (\ref{15}). Besides, as follows from (8b), the
contribution of the second qubit to this state is zero. These
features are presented in Fig.\ref{fig11}. As can be seen from
this figure, the contribution of the second qubit is quite small,
while the damping of the excitation amplitudes of the first and
third qubits is rather slow.

\begin{figure}
  \includegraphics[width=8 cm]{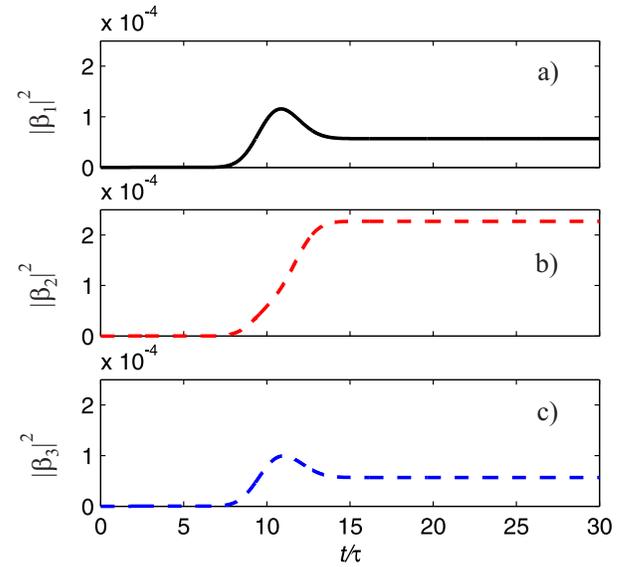}\\
  \caption{The time evolution of the qubits excitation probabilities
   $|\beta_n(t)|^2$, when the probe frequency
  is tuned to the third resonance   $\omega_s=$Re $E_3=1.1\Omega$ ,
  $d=1$ mm, $J/\Omega =0.05$,
   $\Omega/2\pi =5$GHz, $\Gamma/2\pi =10$ MHz, $\Delta /\Omega =0.05
   ( \Omega_2/\Omega =1.05)$.}\label{fig12}
\end{figure}

From Fig.\ref{fig12}, it follows that the amplitudes excitations
of qubits of the third energy level (last line in equation
(\ref{15}) are quite small, although the presence of undamped
subradiant states with almost zero width is seen.

\section{conclusion}
In summary, we have proposed the experimentally achievable method
for the characterization of the collective states of qubits in a
linear chain. The method is based on measuring the time evolution
of the probability of excitation of qubits using a conventional
control pulse technique, which is widely used for recording and
reading out the information in qubit systems. We have examined
this method for a three-qubit linear chain with the nearest
neighbor Ising interaction between qubits. We have shown that the
excitation of qubits by a Gaussian pulse with harmonic filling
allows us to determine the energies, their widths, and the wave
functions of the corresponding collective states. The extension of
this method to more qubits in a chain is straightforward.

\begin{acknowledgments}
Ya. S. G. acknowledges A. N. Sultanov for fruitful discussions.
The work is supported by Ministry of Science and Higher Education
of the Russian Federation under Project
\end{acknowledgments}

%-----------------------------------------------------------
\appendix*
%-----------------------------------------------------------
\section{}
\subsection{Wigner-Weisskopf approximation for the dynamical equations (\ref{19a}, \ref{19b}, \ref{19c})}\label{ap1}

The main assumption is that the quantities $\beta_i{(t)}$ under
integrals in (\ref{19a}, \ref{19b}, \ref{19c}) are slow functions
of time as compared to those in the exponents.Therefore, for times
$t_1\ll  t$ the integrand oscillates very rapidly and there is no
significant contribution to the value of the integral. The most
dominant contribution originates from times $t_1 \approx t$. We
therefore evaluate $\beta_i^{(t)}$ at the actual time $t$ and move
it out of the integrand. In this limit, the decay becomes a
memoryless process (Markov process). To evaluate the remaining
integral in the right hand side of (\ref{A1}) we extend the upper
integration limit to infinity since there is no significant
contribution for $ t_1\gg t$.  Therefore, we obtain:

\begin{equation}\label{A1}
\int\limits_0^t {\beta _i (t')} e^{ - i(\omega _k  - \Omega _i )(t
- t')} dt' \approx \beta _i (t)\int\limits_0^t {} e^{ - i(\omega
_k  - \Omega _i )(t - t')} dt'
\end{equation}

\begin{equation}\label{A2}
\int\limits_0^t {} e^{ - i(\omega _k  - \Omega _i )(t - t')} dt'
\approx \int\limits_0^\infty  {} e^{ - i(\omega _k  - \Omega _i
)(t - t')} dt'
\end{equation}

The last integral is known to be:

\begin{equation}\label{A3}
\int\limits_0^\infty  {} e^{ - i(\omega _k  - \Omega _i )(t - t')}
dt' = \pi \delta (\omega _k  - \Omega _i ) - iP\left(
{\frac{1}{{\omega _k  - \Omega _i }}} \right)
\end{equation}
where P represents the Cauchy principal part, which leads to a
frequency shift. In what follows, we do not write explicitly this
shift, which is assumed to be included in the qubit frequency.
Therefore, the second terms in the equations (\ref{19a},
\ref{19b}, \ref{19c}) can be rewritten as follows:

\begin{widetext}
\begin{equation}\label{A4}
\sum\limits_k {\left| {g_k^{(i)} } \right|^2 \int\limits_0^t
{\beta _i (t')} e^{ - i(\omega _k  - \Omega _i )(t - t')} dt'}  =
\beta _i (t)\pi \sum\limits_k {\left| {g_k^{(i)} } \right|^2
\delta (\omega _k  - \Omega _i )}  \equiv \beta _i
(t)\frac{{\Gamma _i }}{2}
\end{equation}
\end{widetext}

Next, we apply the same procedure to cross terms in (\ref{19a},
\ref{19b}, \ref{19c}).

\begin{equation}\label{A5}
\begin{array}{l}
 \sum\limits_k {g_k^{(i)} g_k^{(j)*} e^{ik(x_i  - x_j )} e^{i(\Omega _i  - \Omega _j )t} \int\limits_0^t {\beta _j (t')} e^{ - i(\omega _k  - \Omega _j )(t - t')} dt'}  \\
  = \sum\limits_k {g_k^{(i)} g_k^{(j)*} e^{ik(x_i  - x_j )} e^{i(\Omega _i  - \Omega _j )t} \beta _j (t)\pi \delta (\omega _k  - \Omega _j )}  \\
 \end{array}
\end{equation}

In (\ref{A4}) the quantity $\Gamma_i$ is the rate of spontaneous
emission for the $i$-th qubit.

\begin{equation}\label{A6}
\Gamma _i  = 2\pi \sum\limits_k {\left| {g_k^{(i)} } \right|^2
\delta (\omega _k  - \Omega _i )}
\end{equation}

In a one dimensional case the summation over $k$ is replaced by
the integration:

\begin{equation}\label{A7}
\sum\limits_k {}  \Rightarrow 2\frac{L}{{2\pi }}\int\limits_{ -
\infty }^\infty  {dk}  = \frac{L}{{2\pi }}4\int\limits_0^\infty
{d\left| k \right|}  = \frac{{2L}}{{\pi \upsilon _g
}}\int\limits_0^\infty  {d\omega _k }
\end{equation}
where the factor of $2$ arises from summing over the two
polarization states associated with each $ k$-vector, and we take
a linear frequency dispersion $\omega_k=v_g|k|$ well above the
cutoff frequency of a waveguide.

Applying the prescription (\ref{A7}) to (\ref{A6}) we express the
coupling $g^{(i)}_{\Omega_i}$ at the qubit resonance frequency
$\Omega_i$ in terms of the rate of spontaneous emission
$\Gamma_i$:

\begin{equation}\label{A8}
\left| {g_{\Omega _i }^{(i)} } \right|^2  = \frac{{\Omega _i D_i
^2 }}{{2\hbar \varepsilon _0 V}} \equiv \frac{{\Gamma _i \upsilon
_g }}{{4L}}
\end{equation}

With the use of (\ref{A7}) and (\ref{A8})  we write the last line
in (\ref{A5})  in the following form:

\begin{equation}\label{A9}
\begin{array}{l}
 \sum\limits_k {g_k^{(i)} g_k^{(j)*} e^{ik(x_i  - x_j )} e^{i(\Omega _i  - \Omega _j )t} \beta _j (t)\pi \delta (\omega _k  - \Omega _j )}
 \\\\
  = \frac{{\sqrt {\Gamma _i \Gamma _j } }}{2}\left( {\frac{{\Omega _j }}{{\Omega _i }}} \right)^{1/2} e^{i\frac{{\Omega _j }}{c}(x_i  - x_j )} e^{i(\Omega _i  - \Omega _j )t} \beta _j (t) \\
 \end{array}
\end{equation}

Now we pay attention to the first terms in the right hand sides of
Eqs. \ref{19a}, \ref{19b}, \ref{19c}. The initial wave packet
$\widetilde{\gamma}_k(0)$ must be normalized to unity:

\begin{equation}\label{A10}
\sum\limits_k {\left| {\tilde \gamma _k (0)} \right|^2 }  =
\frac{\pi }{L}\sum\limits_k {\left| {\gamma _k (0)} \right|^2 }  =
\frac{\pi }{L}2\frac{L}{{2\pi }}\int\limits_{ - \infty }^\infty
{\left| {\gamma _k (0)} \right|^2 dk}  = 1
\end{equation}

The Gaussian envelope $\gamma_k(0)$ defined in (\ref{21})
automatically satisfies this condition:

\begin{widetext}
\begin{equation}\label{A11}
\begin{array}{l}
  - i\sum\limits_k {g_k^{(i)} \tilde \gamma _k (0)e^{ikx_i } e^{ - i(\omega _k  - \Omega _i )t} }  =  - i\sqrt {\frac{{\Gamma _i \upsilon _g }}{{4L}}} \sqrt {\frac{{\omega _s }}{{\Omega _i }}} \frac{L}{{2\pi }}2\sqrt {\frac{\pi }{L}} \int\limits_{ - \infty }^\infty  {dk\gamma _k (0)e^{ikx_i } e^{ - i(\omega _k  - \Omega _i )t} }  \\
  =  - i\sqrt {\frac{{\Gamma _i \upsilon _g }}{{4\pi }}} \sqrt {\frac{{\omega _s }}{{\Omega _i }}} \int\limits_{ - \infty }^\infty  {dk\gamma _k (0)e^{ikx_i } e^{ - i(\omega _k  - \Omega _i )t} }  \\
 \end{array}
\end{equation}
\end{widetext}
where $\omega_s$ is the external excitation frequency.

Integral in (\ref{A11}) of Gaussian envelope can be analytically
calculated which results in the system of linear differential
equations (\ref{22a}, \ref{22b}, \ref{22c}) from the main text.

It is noteworthy that the waveguide length $L$ does not explicitly
enter in these equations. It implicitly enters only in the
definition of $\Gamma_i$ in (\ref{A8}) the value of which is taken
from experiments.

\end{document}